\documentclass[fleqn,twoside]{article}
\usepackage{espcrc2}

\usepackage{graphicx}
\usepackage[figuresright]{rotating}

\newcommand{\AmS}{{\protect\the\textfont2
  A\kern-.1667em\lower.5ex\hbox{M}\kern-.125emS}}

\title{
{
\vspace{-3.35cm} \normalsize \hfill
\parbox{30mm}{DESY 02-124\\Bicocca-FT-02-18\\MS-TP-02-2\\HU-EP-02/37
\\September 2002}
}\\[15mm]
\vspace{-0.15cm}
Recent results on the running coupling in QCD with two massless
flavours\thanks{Talk~given~by~M.~Della~Morte.~Work supported in part by the 
European Community's Human Potential Programme, contract 
HPRN-CT-2000-00145, Hadrons/Lattice QCD.}}
\author{M.~Della Morte\address[Zeu]{DESY Zeuthen, 15738
 Zeuthen, Germany}, R.~Frezzotti\address[Mi]{INFN-Milano and Universit\`a 
Milano--Bicocca, Piazza della Scienza 3, 20126 Milano, Italy},
J.~Heitger\address[Mu]{WWU M\"unster, Institut f\"ur Theoretische Physik,
Wilhelm--Klemm-Str. 9, 48149 M\"unster, Germany},
F.~Knechtli\address[HU]{Institut f\"ur Physik, Humboldt--Universit\"at zu 
Berlin, Invalidenstr. 110, 10115 Berlin, Germany}, J.~Rolf\addressmark[HU],
R.~Sommer\addressmark[Zeu] and U.~Wolff\addressmark[HU] \newline
(ALPHA Collaboration)}

\begin{document}

\begin{abstract}
We report on the latest results on 
the running coupling of two flavour QCD in the Schr\"odinger
functional scheme. Results for the step scaling function are
obtained from simulations on lattices $L/a=8$ and $L/a=16$
which  confirm the first results from
lattices $L/a=4,5,6$ presented one year ago by the ALPHA collaboration.
We  also discuss  some algorithmic aspects, in particular concerning the 
occurrence of metastable states. A modified sampling, in order to estimate 
the proper weight of
these states in the path integral, is proposed and tested.
\vspace{1pc}
\end{abstract}

\maketitle

\section{INTRODUCTION}
The relation between the short distance regime of QCD and its low energy 
sector is a problem which can be addressed by lattice techniques. The $\Lambda$
parameter, characterizing the coupling at large energy, can be related to a 
hadronic quantity by a recursive finite--size technique, avoiding in such a 
way the multiple scale problem. This method has been successfully applied to 
the zero flavour approximation of QCD~\cite{q_coup} and first results
extending it to QCD with two massless flavours have been presented 
last year by our collaboration~\cite{coup_lett}.
 Here we report about more recent  results which allow for a better
 estimate of systematic effects on our determination of the $\Lambda$ 
parameter in the two flavour theory. This result is essential also
in the computation of the running quark mass~\cite{run_mass}.

We consider the Schr\"odinger functional (SF), defined as the Euclidean 
partition function of QCD on a cylinder of size $L^3 \times T$
\vspace{-0.12cm}
\begin{equation}
e^{-\Gamma}=\int_{T\times L^3} D[U,\psi,\bar{\psi}] e^{-S} \;,
\end{equation}
\vspace{-0.12cm}
with periodic (up to a phase) boundary conditions in the spatial directions
and Dirichlet boundary conditions at $x_0=0,T$.
The spatial links at $x_0=0,T$ are fixed to diagonal SU(3) matrices
specified in terms of $L$ and one angle $\eta$ while the 
quark fields are given by Grassmann values $\rho$, $\bar{\rho}$ and
$\rho'$, $\bar{\rho}'$, which are used as sources and then set to zero.

We work with an O($a$) improved Wilson action and improved operators, employing
 the non--perturbative  $c_{\rm SW}$~\cite{csw} coefficient and one and 
two--loop values for the
 boundary coefficients $\tilde{c}_t$ and $c_t$~\cite{cA,ctctt}. The 
improvement coefficient $c_{\rm A}$ is also set to its one--loop value. It 
enters our definition of quark mass $m$ via the PCAC relation.
The mass is 
required to be consistent with zero~\cite{coup_lett}.

For $T=L$, the SF coupling $\bar{g}^2$ is defined~\cite{q_coup,SFYM}
\vspace{-0.07cm}
\begin{equation}
\left. (\partial\Gamma/\partial \eta) \right|_{\eta=0} = 
\left. \langle \partial S/\partial \eta \rangle \right|_{\eta=0} =
k\bar{g}^{-2}(L)
\end{equation}
\vspace{-0.07cm}
with $k$ fixed by the condition $\bar{g}^2=g_0^2+O(g_0^4)$.
Here $L$ plays the role of a renormalization scale.
\vspace{-0.1cm}
\section{NUMERICAL RESULTS}
\vspace{-0.1cm}
The main quantity in the present computation is the step scaling function (SSF)
\vspace{-0.02cm}
\begin{equation}
\sigma(u)=\bar{g}^2(2L) |_{\bar{g}^2(L)=u,m=0} \; ,
\end{equation}
which can be viewed as a non--perturbative integrated form of the 
$\beta$--function. Once $\sigma(u)$ has been computed in the continuum limit 
and for a large enough range of values, the equation
\begin{equation}
\sigma(\bar{g}^2(L/2))= \bar{g}^2(L) 
\end{equation}
can be recursively solved $n$ times (for explicit equations see 
e.g.~\cite{run_mass}) starting from the hadronic 
scale $L_{\rm max}$ with $\bar{g}^2(L_{\rm max})=u_{\rm max}$.
 In this way we obtain
values for $\bar{g}^2(2^{-n}L_{\rm max})$. For large $n$ this coupling is
perturbative and we compute the $\Lambda$ parameter in units of $L_{\rm max}$
making use of the three--loop $\beta$ function in the SF scheme~\cite{coup_lett,ctctt}
\begin{eqnarray}
&&\Lambda L_{\rm max} =  2^n (b_0\bar{g}^2)^{-b_1/2b_0^2} 
\exp\left\{ -\frac1{2b_0\bar{g}^2}  \right\} \nonumber\\
&&\times
\exp\left\{ -\int_0^{\bar{g}} dx \left[        
\frac1{\beta(x)} + \frac1{b_0 x^3} -\frac{b_1}{b_0^2 x}  \right]\right\}\
\label{Lambdapert}
\end{eqnarray}
Later we will have to relate $L_{\rm max}$ to a physical scale, e.g. by
determining $L_{\rm max}F_{\pi}$.
%
\begin{table}[htbp]\label{table:1}
\vspace{-0.5cm}
\small{
\begin{center}
  \begin{tabular}{cllll} \hline
  $\!\!\!\!L/a\!\!\!\!$ & \multicolumn{1}{c}{$\!\!u$}  & 
         \multicolumn{1}{c}{$\!\!\Sigma$} 
    & \multicolumn{1}{c}{$\!\!u$}  & \multicolumn{1}{c}{$\!\!\Sigma$} \\ \hline
          &               &               \\[-2ex]
    $\!\!\!\!8\!\!\!\!$    & $\!\!0.9807(17)$ & $\!\!1.0745(55) $
          &$\!\!1.508(4)$& $\!\!1.716(14)$\\[0.5ex]  
    $\!\!\!\!8\!\!\!\!$    & $\!\!1.1818(29)$ & $\!\!1.3338(58) $
          & $\!\!2.014(10)$   &  $\!\!2.475(31) $   \\[0.5ex]
    $\!\!\!\!8\!\!\!\!$    &           &               
          & $\!\!2.479(13)$  &$\!\!3.348(48)$\\[0.5ex]
    $\!\!\!\!4\!\!\!\!$    
                             &                    &
          & $\!\!3.334(11)$    & $\!\!5.513(42) $  \\
    $\!\!\!\!6\!\!\!\!$    
          &                    &
          & $\!\!3.326(20)$    & $\!\!5.62(9)  $  \\[0.5ex]\hline
\end{tabular}
\end{center}
\caption{New data for the lattice SSF 
$\Sigma(u,a/L)$. $c_{\rm t} (g_0)$ was set
to its one--loop value for the left part, and to two--loop for  the two 
rightmost columns.}
\vspace{-0.65cm}
}
\end{table}
Our new simulations and results 
are summarized in Table~1.
These allow for  more reliable
continuum limit extrapolations of the SSF. 
%

\begin{figure}[htb]
\vspace{-0.3cm}
\includegraphics[width=18pc]{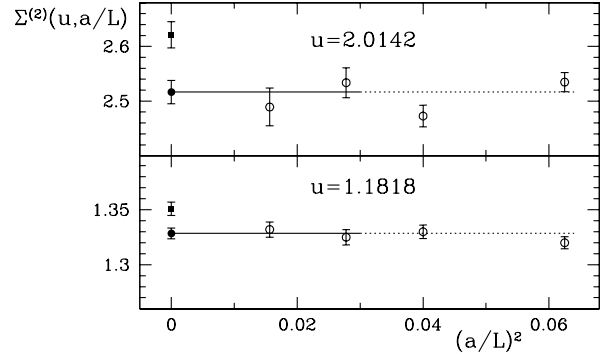}
\vspace{-1.32cm}
\caption{Scaling plot for $\Sigma^{(2)}$. 
Black squares indicate the continuum limit  extrapolation for $N_{\rm f}=0$.}
\label{fig_scal}
\vspace{-0.7cm}
\end{figure}

\begin{figure}[htb]
\vspace{-2.2cm}
\includegraphics[width=19pc]{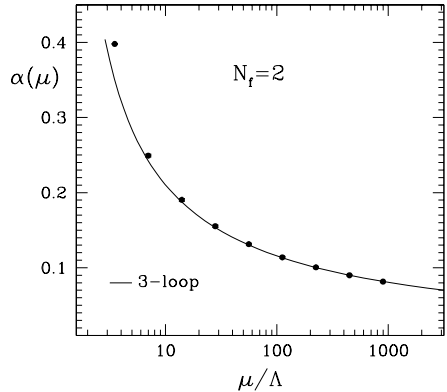}
\vspace{-2.4cm}
\label{run_fig}
\caption{Running of $\alpha\!=\!\bar{g}^2/4\pi$ in the SF scheme.} 
\vspace{-0.7cm}
\end{figure}

We extract the continuum limit values for the SSF using data for the two--loop
improved observable $\Sigma^{(2)}(u,a/L)$~\cite{coup_lett}  from 
$L/a=6$ and $L/a=8$.
The scaling of $\Sigma^{(2)}$ vs. $(a/L)^2$ is shown in Figure~\ref{fig_scal} 
for two different values of $u$. For comparison we included in the plot the 
continuum results in the quenched case (black squares) for the same
value of the couplings.
The function $\sigma(u)$ is obtained interpolating the continuum 
limit estimates
by a sixth--order polynomial with the first three coefficients constrained by
 perturbation theory. Starting from $u_{\rm max}\!=\!3.3$ or $u_{\rm max}\!=\!5$ we 
can then estimate the quantity $\Lambda L_{\rm max}$ as discussed above. 
 Results are summarized in Table~2. 
\begin{table}[htbp]
\vspace{-0.4cm}
\begin{center}
\begin{tabular}
{ccc|ccc}
\hline
  & \multicolumn{2}{c|}{$\bar{g}^2(L_{\rm max})=3.3$} & 
  & \multicolumn{2}{c}{$\bar{g}^2(L_{\rm max}')=5$} \\
$n$ & c.l.  6/8 & $L/a=5$ & $n$ &  $\!\!6/8$ & $\!\!L/a=5$ \\ 
\hline
 5 &  1.82(5) &  1.87 &  6 &  1.23(5) &  1.27 \\
 6 &  1.84(6) &  1.89 &  7 &  1.25(6) &  1.28 \\
 7 &  1.85(7) &  1.91 &  8 &  1.26(7) &  1.30 \\[0.5ex]
\hline
\end{tabular}
\vspace{0.15cm}
\caption{Values for $-\ln(\Lambda L_{\rm max})$ for two choices of 
$L_{\rm max}$.}
\end{center}
\vspace{-1cm}
\end{table}
We use the difference between continuum values and data from $L/a=5$ 
to estimate our systematic uncertainty, which we linearly add to the
statistical errors. This yields
\begin{eqnarray}
\ln(\Lambda L_{\rm max}) &=& -1.85(13) \quad [\bar{g}^2(L_{\rm max})=3.3] 
\nonumber \\ 
\ln(\Lambda L_{\rm max}') &=& -1.26(11) \quad [\bar{g}^2(L_{\rm max}')=5] \, .
\label{Lam5}
\end{eqnarray}
Comparing with the results in~\cite{coup_lett}, we see that the systematic 
uncertainty on $\ln(\Lambda L_{\rm max})$ is nearly halved, while the central
values are fully consistent. 
The non--perturbative evolution
of $\alpha(\mu)=\bar{g}^2(L)/4\pi$ $(\mu=1/L)$ starting from $\bar{g}^2=5$
is plotted in Figure~\ref{run_fig}.
Statistical and systematical errors (difference between continuum limit and 
$L/a=5$) are too small to be visible in the plot. 
The figure shows that differences between Monte
Carlo results and three--loop perturbation theory become appreciable for large
 values of the coupling, while for high energies the perturbative 
approximation very closely reproduces the non--perturbative results.
\vspace{-0.15cm}
\section{METASTABLE STATES}
\vspace{-0.05cm}
For the largest values of the coupling in our study ($\bar{g}^2\simeq 3.3$ and
$\bar{g}^2 \simeq 5.5$) we observed a long tail towards negative values in the
distribution of
$\partial S/\partial \eta$. In addition, measuring by a cooling procedure
the pure gauge contribution to  the action $S_{\rm G}^{cool}$, 
and to the coupling $\partial S_{\rm G}^{cool} / \partial \eta $,
we observed occurrences of
metastable states in our Monte Carlo histories. The action $S_{\rm G}^{cool}$
for these states is consistent with the value for a secondary solution of the
field equations~\cite{SFYM}, given our choice for the boundary fields.
This solution appears to be a local minimum by numerical evidence.

In order to properly estimate the weight of these states
we enhanced their occurrence through a modified sampling~\cite{q_coup},
adding to the 
HMC effective action a term
\begin{equation}
\left. \gamma{{\partial S_{\rm G}}\over{\partial \eta}} \right|_{\eta=0} + 
{{1}\over{w_{\gamma}}}(\gamma-\gamma_0)^2 \;,
\end{equation}
where $\gamma_0$ and $w_{\gamma}$ are fixed values, while $\gamma$ is a 
dynamical variable. The expectation values in the 
original ensemble are then obtained by reweighting properly the observables.

Defining a quantity $q$ whose value is 1 for metastable states and 0 
otherwise (so that $\langle \delta_{q1}\rangle +\langle \delta_{q0}\rangle=1$),
for an observable $O$ one can write
\begin{eqnarray}
\langle O\rangle &=& \langle \delta_{q1}O\rangle + 
\langle\delta_{q0}O\rangle
\\ \nonumber 
\vspace{0.15cm}
 &=&\langle \delta_{q1}O\rangle + \langle O \rangle_{\bar{1}} 
(1-\langle \delta_{q1}\rangle) \; ,
\end{eqnarray}
where $\langle O \rangle_{\bar{1}}=\langle \delta_{q0}O\rangle/
\langle \delta_{q0}\rangle$. If the main contribution to 
$\langle O \rangle$ comes from the $q=0$
sector then  a precise estimate of $\langle O \rangle$ just requires
a precise estimate of $\langle O \rangle_{\bar{1}}$, which can be 
obtained by an  algorithm which  samples only the 
$q=0$ sector, together with  rough estimates  of 
$\langle \delta_{q1}\rangle$  
and  $\langle \delta_{q1}O\rangle$ which can be 
obtained by the modified sampling.

By a set of simulations on $L/a=8$ and $12$ using this approach we 
could estimate the effect of metastable states on $\bar{g}^2$ to be nearly
$0.10(2)\%$ for $\bar{g}^2 \simeq 3.3$.
This result is independent, within errors, from $L/a$.
Moreover, for this value of the coupling, we noticed a slightly better 
behavior for the PHMC algorithm in comparison with HMC (smaller values of 
$\gamma_0$ needed in order to generate a certain number of transitions
from/to a metastable state). 

For $\bar{g}^2\simeq 5.5$, the effect of metastable states  is
much larger, as expected, resulting in a $4$ to $5\%$ contribution.
However their sampling is also enhanced in this case using both PHMC or HMC
already in the original ensemble.
Indeed we repeated the $L/a=12$ simulation for $\bar{g}^2 \simeq 5.5$ using
ordinary PHMC as in Ref.~\cite{coup_lett} but disentangling the occurrence of
 metastable states. This turned out to be around $6\%$ and the result
for $\bar{g}^2$ was fully consistent with the number in Ref.~\cite{coup_lett}. 

{\bf Acknowledgments}. We thank NIC/DESY Zeuthen for allocating computer time 
on the
APEmille machines to this project, and the APE group for their support.

\end{document}